\pdfoutput=1

\documentclass[reprint,amsmath,amssymb,floatfix,aps,longbibliography]{revtex4-1}
\usepackage{graphics,graphicx,float}

\usepackage[pdftex,
            pdfauthor={Steven A. Frank},
            pdftitle={},
            pdfsubject={Natural selection. II. Developmental variability and evolutionary rate}]{hyperref} 

\def\citeyear{\citep}
\def\autocite{\citep}

\newcommand{\xbar}{\bar{x}}
\newcommand{\N}{\mathcal{N}}
\newcommand{\dd}{{\hbox{\rm d}}}

\newcommand{\Gg}{\gamma}

\newcommand{\Gs}{\sigma}

\newcommand{\Eq}[1]{Eq.~(\ref{eq:#1})}

\newcommand{\Fig}[1]{Fig.~\ref{fig:#1}}

\newcommand{\boldrule}{\hrule height 1.2pt}
\newcommand{\noterule}{\medskip\boldrule\medskip}	% for notes

\newcount\BoxNum \BoxNum 1\relax
\makeatletter
\newcommand{\boxlabel}[1]{%
  \protected@write \@auxout {}{\string \newlabel {box:#1}{{\the\BoxNum}{\thepage}{\noexpand\relax}%
  	{\@ifundefined{hyper@@anchor}{\relax}{box.\the\BoxNum}}%
  	{}}}%
  \@ifundefined{hyper@@anchor}{}{\hypertarget{box.\the\BoxNum}{}}%
  \advance\BoxNum 1\relax}
\makeatother
\newcommand{\Boxx}[1]{Box~\ref{box:#1}}
\newcommand{\BoxLabel}{Box~\the\BoxNum}

\begin{document}

\title{Natural selection. II. Developmental variability and evolutionary rate}

\author{Steven A.\ Frank}
\email[email: ]{safrank@uci.edu}
\homepage[homepage: ]{http://stevefrank.org}
\affiliation{Department of Ecology and Evolutionary Biology, University of California, Irvine, CA 92697--2525  USA}

\begin{abstract}

In classical evolutionary theory, genetic variation provides the source of heritable phenotypic variation on which natural selection acts.  Against this classical view, several theories have emphasized that developmental variability and learning enhance nonheritable phenotypic variation, which in turn can accelerate evolutionary response. In this paper, I show how developmental variability alters evolutionary dynamics by smoothing the landscape that relates genotype to fitness.  In a fitness landscape with multiple peaks and valleys, developmental variability can smooth the landscape to provide a directly increasing path of fitness to the highest peak.  Developmental variability also allows initial survival of a genotype in response to novel or extreme environmental challenge, providing an opportunity for subsequent adaptation.  This initial survival advantage arises from the way in which developmental variability smooths and broadens the fitness landscape.  Ultimately, the synergism between developmental processes and genetic variation sets evolutionary rate\footnote{\href{http://dx.doi.org/10.1111/j.1420-9101.2011.02373.x}{doi:\ 10.1111/j.1420-9101.2011.02373.x} in \textit{J. Evol. Biol.}}\footnote{Part of the Topics in Natural Selection series. See \Boxx{preface}.}.

\end{abstract}

\maketitle

\begin{quote}
{\small In evolutionary biology, environmentally induced modifications come under unfinished business $\ldots$ There have been repeated assertions of both their importance and their triviality, a lot of discussion with no consensus.~$\ldots$ Yet the debate has continued over such concepts as genetic assimilation, the Baldwin effect, organic selection, morphoses, and somatic modifications. So much controversy over the span of a century suggests that a problem of major significance remains unsolved} \autocite[p.~498]{west-eberhard03developmental}.
\end{quote}

\section{Introduction}

A single genotype produces different phenotypes. Developmental programs match the phenotype to different environments. Intrinsic developmental fluctuations spread the distribution of phenotypes.  Extrinsic environmental fluctuations perturb developmental trajectory. These nonheritable types of phenotypic variation are common.

Nonheritable phenotypic variation is not transmitted through time.  Thus, nonheritable variation would seem to be irrelevant for evolutionary change, which instead depends on the genetic component of variation.  However, nonheritable phenotypic variation can, in principle, affect evolutionary rate.  At first glance, that contribution of nonheritable phenotypic variation to evolutionary rate appears to be a paradox. 

Many different theories, commentaries, and controversies turn on this paradox (\Boxx{literature}). The literature has followed a consistent pattern.  Detailed theories relate developmental variability to accelerated evolution.  Counterarguments ensue.  Listings of complicated examples claim to support the theory. Refinements to the theory develop.  

In the end, few compelling examples relate nonheritable phenotypic variability to evolutionary rate.  The literature is hard to read. Enthusiasts extend the concepts and keep the problem alive.  Through the enthusiasts' promotions, many have heard of the theory.  But, in practice, few consider the role of nonheritable phenotypic variability in their own analyses of evolutionary rate.  Almost everyone ignores the problem.  

In this article, I emphasize simple theory that relates nonheritable phenotypic variability to evolutionary rate.  Understanding the paradoxical relation between nonheritable phenotypic variability and evolutionary rate is an essential step in reasoning about many evolutionary problems. 

This article is primarily a concise tutorial to the basic concepts (see \Boxx{preface}).  I briefly mention some of the history (\Boxx{literature}) and recent, more advanced literature (\Boxx{literature2}).

\section{Smoothing the evolutionary path}

The distribution of phenotypes for a given genotype is called the \textit{reaction norm.}  All theories come down to the fact that a broad reaction norm smooths the path of increasing fitness.  Once one grasps the smoothing process, many apparently different theories become easy to understand.

The next section gives the mathematical expression for the smoothing of fitness by the reaction norm.  \Fig{smooth} explains the mathematics with a simple example.  

\subsection{The reaction norm smooths fitness}

We need to track three quantities.  First, fitness, $f(x)$, varies according to the particular phenotype expressed, $x$.  

\begin{figure}[H]
\begin{minipage}{\hsize}
\parindent=15pt
\noterule
{\bf \noindent\BoxLabel. Topics in the theory of natural selection}
\noterule
This article is part of a series on natural selection.  Although the theory of natural selection is simple, it remains endlessly contentious and difficult to apply.  My goal is to make more accessible the concepts that are so important, yet either mostly unknown or widely misunderstood.  I write in a nontechnical style, showing the key equations and results rather than providing full derivations or discussions of mathematical problems.  Boxes list technical issues and brief summaries of the literature.   
\noterule
\end{minipage}
\end{figure}
\boxlabel{preface}

Second, the phenotype expressed varies according to the reaction norm.  Read $p(x|\xbar)$ as the probability of expressing the phenotype $x$ given a genotype with average phenotype $\xbar$. 

Third, we must calculate $F(\xbar)$, the expected fitness for a genotype with average phenotype $\xbar$.   We obtain the expected fitness by summing up the probability, $p$, of expressing each phenotype multiplied by the fitness, $f$, of each phenotype.  That sum is 
\begin{equation}\label{eq:discrete}
  F(\xbar)=\sum p(x|\xbar)f(x),
\end{equation}
taken over all the different phenotypes, $x$.  We often measure $x$ as a continuous variable.  The sum is then equivalently written as
\begin{equation}\label{eq:cont}
  F(\xbar)=\int p(x|\xbar)f(x)\dd x.
\end{equation}
This equation shows how one averages the fitness, $f(x)$, for each phenotypic value, $x$, over the reaction norm, $p(x|\xbar)$, to obtain the expected fitness of a genotype, $F(\xbar)$. We label each genotype by its average phenotype, $\xbar$.  The expected fitness of a genotype, $F(\xbar)$, is what matters for evolutionary process \autocite{frank11natural}.  

The averaging of expected fitness over the reaction norm is the key to the entire subject.  Averaging over the reaction norm, $p$, flattens and smooths the fitness function, $f$.  This smoothing makes the curve for expected fitness, $F$, have lower peaks and shallower valleys than the original fitness curve, $f$.  The smoothing of $F$ changes evolutionary dynamics.  The whole problem comes down to understanding how reaction norms smooth fitness, and the consequences of a smoother relation between genotype and fitness.

\subsection{Example of continuous smoothing}

\begin{figure*}[t]
\includegraphics[width=0.75\hsize]{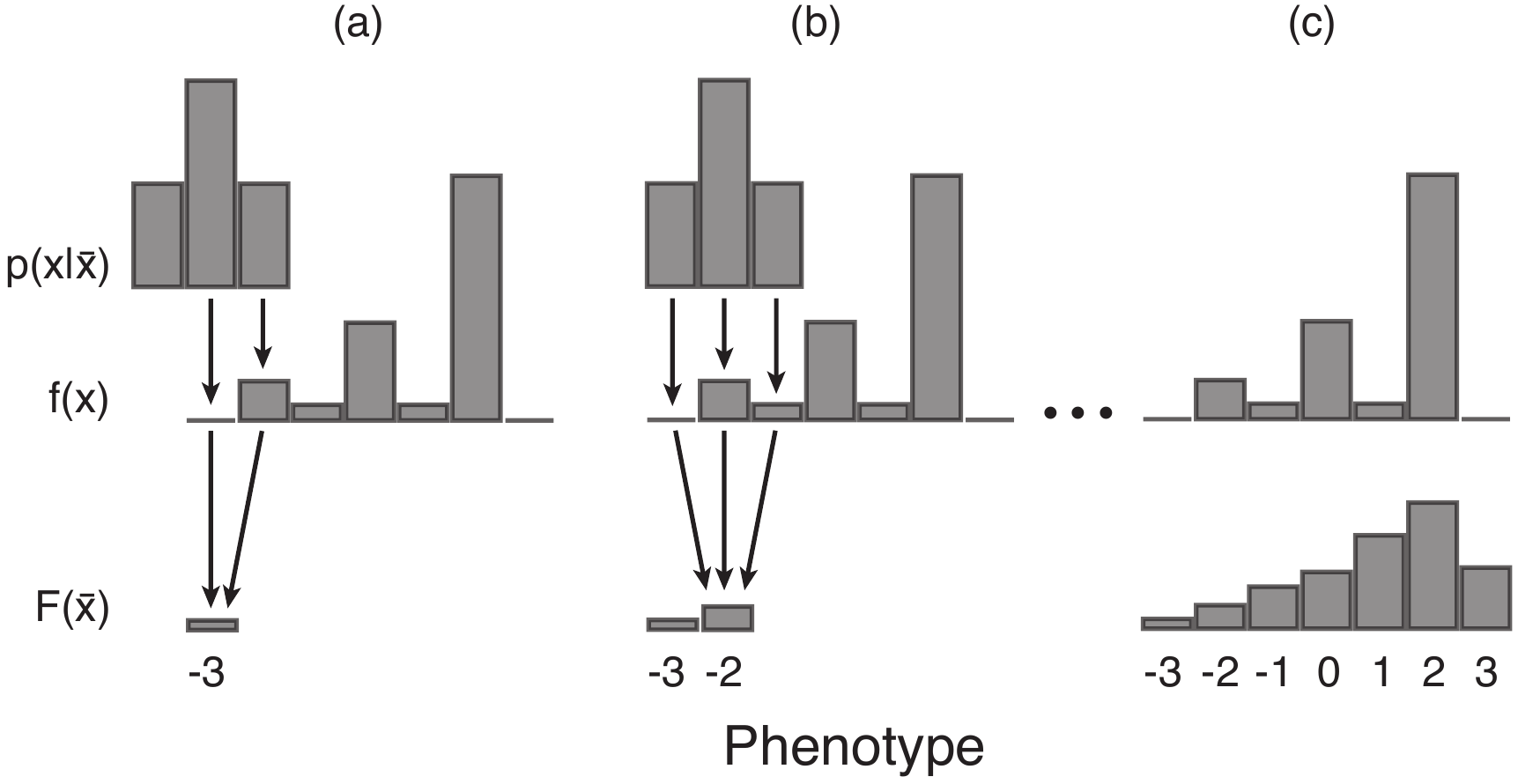}
\caption{The reaction norm smooths the fitness landscape.  This simple example illustrates the calculation of the expected fitness for each genotype, following \Eq{discrete}.  (a) The calculation of expected fitness, $F(\xbar)$, for the smallest average phenotype, $\xbar=-3$.  For that average phenotype, the reaction norm, $p(x|\xbar)$, shows the probabilities of expressing different phenotypes, $x$.  In this case, the peak of the reaction norm matches the average value, and each phenotype $\pm 1$ occurs half as often as the peak value.  To get the expected fitness for a reaction norm centered at $\xbar=-3$, one sums up the probability $p(x|\xbar)$ for each phenotype, $x$, multiplied by the fitness for each phenotype, $f(x)$.  The arrows illustrate the summation.  (b) The expected fitness, $F(\xbar)$, for each increase in $\xbar$, is calculated by the same summation process, shifting the reaction norm to the right by one to get the proper value for each $\xbar$. (c) The full transformation is shown between the fitness for each phenotypic value, $f(x)$, and the expected fitness, $F(\xbar)$, for each genotype with reaction norm $p(x|\xbar)$ and average phenotype $\xbar$.  The reaction norm smooths the multipeaked fitness function, $f(x)$, into the single-peaked fitness function $F(\xbar)$.  Evolutionary dynamics depend on genotypic fitnesses, $F$.  Thus, the reaction norm transforms fitness into a smooth function that allows a direct increasing path to the fitness peak from any starting value for average phenotype.}
\label{fig:smooth}
\end{figure*}

\begin{figure}[t]
\includegraphics[width=0.8\hsize]{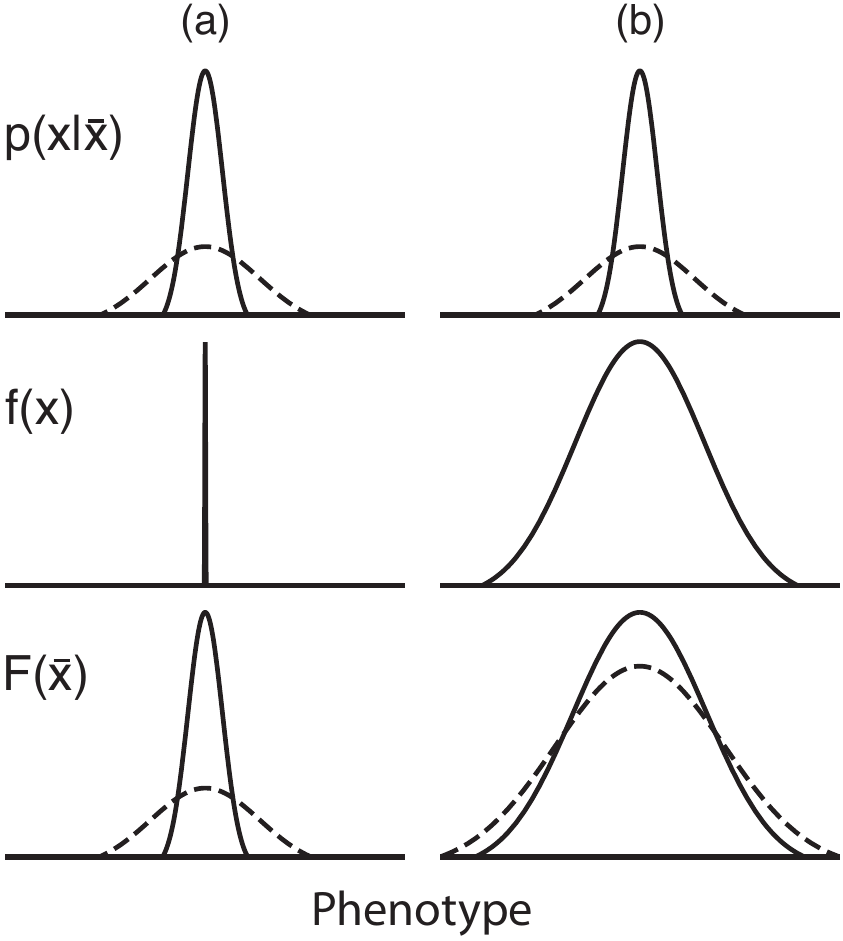}
\caption{Reaction norms and fitness for continuous phenotypes.  Each column shows how the reaction norm, $p(x|\xbar)$, smooths the fitness function, $f(x)$, to give the expected fitness, $F(\xbar)$, for a genotype with average phenotype $\xbar$.  The smoothing follows \Eq{cont}.  These examples use normal distributions that lead to \Eq{contSmooth}.  (a) The solid and dashed reaction norms follow $\N(\xbar,1/2)$ and $\N(\xbar,5)$, respectively. Fitness, $f(x)$, has the shape of a normal distribution with vanishingly small variance, $\N(0,\Gs^2\rightarrow0)$.  Thus, expected fitness, $F(\xbar)$, is the same as the reaction norm. (b) The same structure as in (a), except that $f(x)$ is much wider, following $\N(0,7)$.  Thus, $F(\xbar)$ now has curves $\N(0,7.5)$ and $\N(0,12)$ for solid and dashed curves, respectively.  In each plot, the baseline is set to $4.3\%$ of the peak in that plot.  The baseline truncates phenotypes with low vigor, setting their fitnesses to zero.}
\label{fig:transform}
\end{figure}

\Fig{smooth} shows an example of smoothing with discrete distributions.  It will often be convenient to consider smoothing of continuous variables.  \Fig{transform} shows an example.  The following expressions describe the underlying mathematics.

\begin{figure}[H]
\begin{minipage}{\hsize}
\parindent=15pt
\noterule
{\bf \noindent\BoxLabel. Historical overview}
\noterule
\textcite{schlichting98phenotypic} and \textcite{west-eberhard03developmental} thoroughly review the subject.  Here, I highlight a few key points in relation to this article.  I treat learning and developmental plasticity as roughly the same with regard to potential consequences for evolutionary rate, although one could certainly choose to focus on meaningful distinctions.

In my own reading during the 1980s, I had found the relation between learning and evolutionary rate intriguing but confusing.  Baldwin's \autocite{baldwin96a-new-factor} idea that learning can accelerate evolutionary rate seemed attractive.  \textcite{mayr1982growth}, in his monumental review of biological thought, also discussed various ways in which behavior or flexible developmental programs might alter evolutionary dynamics.  Those ideas seemed potentially important, but it was not easy to grasp the essence.  The literature at that time was not helpful, with a lot of jargon and sometimes almost mystical commentary mixed in with intriguing and creative ideas.

It was clear that learning could slow evolutionary rate.  Different genotypes could, through learning,  end with the same phenotype.  Reducing the phenotypic distinction between different genotypes would generally slow evolutionary rate.  The more intriguing problem concerns the origin of evolutionary novelty or the response to novel or extreme environmental challenge.  Environmental novelty and acceleration of evolutionary response were the primary concern of \textcite{baldwin96a-new-factor}, \textcite{waddington42canalization,waddington53genetic}, and \textcite{west-eberhard03developmental}.  My article also focuses on acceleration of evolutionary response.  

\textcite{hinton87how-learning} clarified the subject with their simple conclusion that:
\begin{quote}
Learning alters the shape of the search space in which evolution operates and thereby provides good evolutionary paths towards sets of co-adapted alleles. We demonstrate that this effect allows learning organisms to evolve \emph{much\/} faster than their nonlearning equivalents, even though the characteristics acquired by the phenotype are not communicated to the genotype.
\end{quote} 
During the past few decades, the fundamental role of smoothed fitness surfaces in biology has not always been recognized as fully as it should be, in spite of several fine papers along that line (see \Boxx{literature2}).  Interestingly, certain computer optimization algorithms take advantage of the increased search speed provided by a process similar to smoothed fitness landscapes \autocite{kirkpatrick83optimization,geyer95annealing}.
\noterule
\end{minipage}
\end{figure}
\boxlabel{literature}

In \Fig{transform}, the reaction norm follows a normal distribution.  In symbols, we write 
\begin{equation*}
  p(x|\xbar) \sim \N(\xbar,\Gg^2),
\end{equation*}
which we read as the probability, $p$, of a phenotype, $x$, for a reaction norm centered at $\xbar$, follows a normal distribution with mean $\xbar$ and variance $\Gg^2$.

For fitness, we write in symbols 
\begin{equation*}
  f(x) \sim \N(0,\Gs^2),
\end{equation*}
which we read as the fitness, $f$, of a phenotype, $x$, has the shape of a normal distribution with mean $0$ and variance $\Gs^2$.  In this case, we assume the center of the fitness distribution is at a phenotypic value of zero to give a fixed point for comparison---any value to center fitness could be used.  The important issue is that fitness falls off from its peak by the pattern of a normal distribution.  The width of the fitness function is set by the variance parameter, $\Gs^2$.

We can now use \Eq{cont} to calculate the expected fitness of a genotype with average phenotype $\xbar$, yielding
\begin{equation}\label{eq:contSmooth}
  F(\xbar) \sim \N(0,\Gg^2+\Gs^2).
\end{equation}
This equation shows that smoothing by the reaction norm, $p$, flattens and widens the shape of the fitness function by increasing the variance of the expression for $F$.  

\subsection{Evolutionary response to novel or extreme challenge}

If a genotype expresses an average phenotype close to the maximum fitness, then a narrow reaction norm has higher fitness than a broad reaction norm.  The lower plots of $F(\xbar)$ in \Fig{transform} illustrate contrasting widths of reaction norms.  Near the peak, the average phenotype closely matches the optimum, and the narrower reaction norm has higher fitness.  This advantage occurs because a narrow reaction norm expresses fewer phenotypes in the tails, away from the optimum.

\begin{figure}[t]
\includegraphics[width=0.8\hsize]{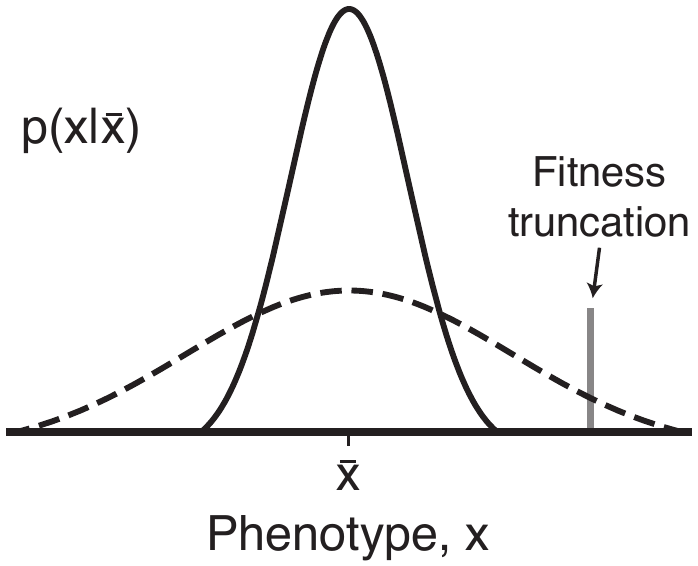}
\caption{Novel environmental challenge or intense competition favors a broad reaction norm.  In this example, both the broad and narrow reaction norms are centered at $\xbar$.  Phenotypes above the truncation point survive. Phenotypes below the truncation point die. None of the phenotypes for the narrow reaction norm are above the truncation point, so all die.  Some of the phenotypes of the broad reaction norm survive.  Those surviving phenotypes may evolve so that their average phenotype, $\xbar$, moves toward the truncation point, improving fitness over time.  Improvement occurs if there is genetic variation for the average phenotype, $\xbar$, of the reaction norm.}
\label{fig:truncate}
\end{figure}

For genotypes with an average phenotype far from the maximum fitness, a broad reaction norm has higher fitness than a narrow reaction norm.  \Fig{truncate} illustrates this advantage for broad reaction norms.  In that figure, both reaction norms are centered at $\xbar$.  Only those phenotypes above the fitness truncation point survive.  The broad reaction norm produces some individuals with phenotypes above the truncation point, whereas the narrow reaction norm has zero fitness.  

If the environment poses a novel or extreme challenge, the broad reaction norm wins.  By contrast, in a stable environment for which the current average phenotype is close to the fitness optimum, the narrow reaction norm wins.  Thus, extreme or novel environmental challenges or intense competition favor a broad reaction norm.  

\textcite{haldane32the-causes} made a similar point when he said: ``Intense competition favors variable response to the environment rather than high average response. Were this not so, I expect that the world would be much duller than is actually the case.''  Holland's \autocite{holland75adaptation} emphasis on exploration versus exploitation is perhaps closer to the problem here.  Broad reaction norms are favored when exploration of novel challenges dominates, whereas narrow reaction norms are favored when exploitation dominates.  Fluctuating environments may also favor a broad reaction norm to increase the chance of matching whatever is favored at any time \autocite{frank11natural}.  Here, I focus on constant challenges to extreme or novel environments.

\subsection{Smoothly increasing fitness path in a multipeak fitness landscape}

\begin{figure}[t]
\includegraphics[width=0.7\hsize]{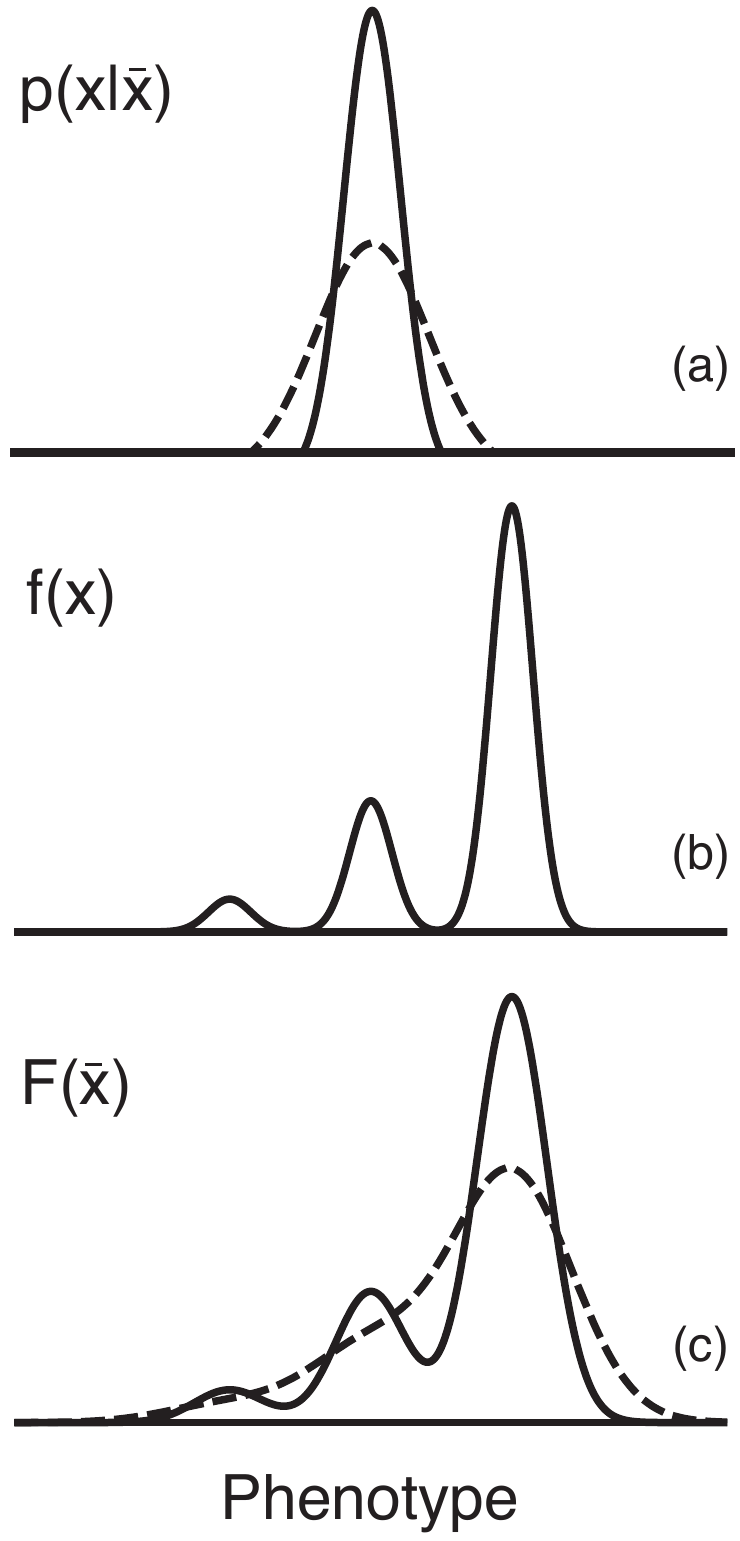}
\caption{A broad reaction norm smooths a multipeak fitness landscape.  (a) The dashed curve shows the broader reaction norm, $p(x|\xbar)$.  (b) The fitness landscape for each particular phenotype, $f(x)$, has multiple peaks.  (c) The broad reaction norm smooths the fitness landscape to a single peak for the relation between the average phenotype for a genotype, $\xbar$, and fitness, $F(\xbar)$.  In this example, the narrow and broad reaction norms follow $\N(0,\Gg^2)$ distributions with variances of $0.04$ and $0.16$, respectively.  Fitness is given by $f(x)=\sum_{i=-1}^1 (3|1+i|^2+1)\N(i,\Gs^2)$, with $\Gs^2=0.0225$. The value of $F(\xbar)$ is calculated from \Eq{cont}, yielding the expression for $f(x)$ in the prior sentence with the variance replaced by $\Gs^2+\Gg^2$. The baseline truncates small values.}
\label{fig:multipeak}
\end{figure}

Much discussion in evolutionary theory concerns how populations shift from a lower fitness peak to a higher fitness peak \autocite{coyne97a-critique}.  For example, in the fitness landscape, $f(x)$, of \Fig{multipeak}b, a population starting on a lower peak must evolve through a valley of lower fitness in order to follow an increasing path to a higher fitness peak.  Natural selection typically follows a path of increasing fitness, so a population may be trapped on a lower peak. 

Most evolutionary analyses use a fitness landscape that relates phenotype, $x$, to fitness, $f(x)$.  However, the proper measure should relate the average phenotype of a genotype, $\xbar$, to the expected fitness, $F(\xbar)$ \autocite{frank11natural}.

A sufficiently broad reaction norm smooths a multipeak fitness landscape, $f(x)$, into a smooth landscape, $F(\xbar)$, with a single peak (\Fig{multipeak}c).  A broad reaction norm will typically perform badly near a fitness peak, but allow much more rapid evolutionary advance to a higher fitness peak.  Once again, we see that broad reaction norms exploit current fitness opportunities relatively poorly but gain by enhanced exploration and achievement of novel adaptations.

\section{Dimensionality and discovery}

The reaction norm may be generated randomly by perturbations in development.  If so, then exploration of the fitness landscape by a broad reaction norm is a type of random search.  Figs.~\ref{fig:truncate} and \ref{fig:multipeak} show that random search can greatly increase the rate of adaptation, particularly to novel environmental challenges.

Those previous examples showed the reaction norm and fitness both varying across a single dimension.  A broad reaction norm spreads phenotypes along that single dimension, increasing the chance that some individuals will have high fitness.

Now consider the much more difficult search problem that arises in higher dimensions \autocite{gavrilets04fitness}.  Suppose, for example, that adapting to a novel environmental challenge requires multiple phenotypic changes to work together in a harmonious way.  Think of each particular phenotypic change as a trait in its own dimension, so that the search now occurs in multiple dimensions.  If the reaction norm simply generates random phenotypes in each dimension, then there is little chance of getting simultaneous matching phenotypes in multiple dimensions.  

\begin{figure}[t]
\includegraphics[width=0.7\hsize]{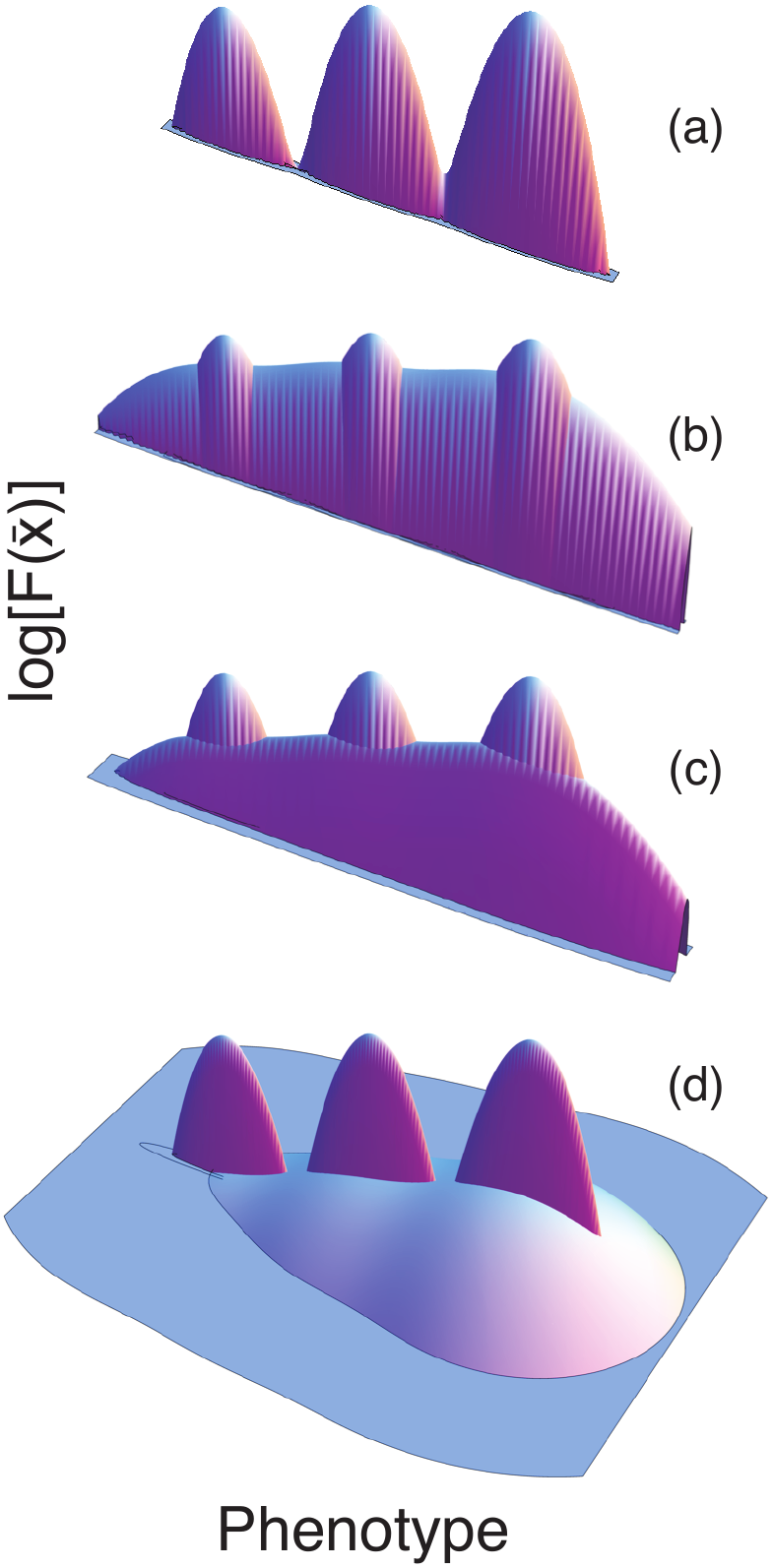}
\caption{A broad reaction norm performs poorly when fitness is concentrated in a lower dimension.  (a) The bivariate analogy of the fitness landscape in \Fig{multipeak}b, scaled logarithmically.  The primary dimension has variance $\Gs^2_1=0.0225$ corresponding to standard deviation $\Gs_1=0.15$, as in \Fig{multipeak}b.  The secondary (narrow) dimension has standard deviation $\Gs_2=0.1\Gs_1$.  (b) Fitness landscape smoothed by a reaction norm concentrated in the same dimension as fitness. The variance of the reaction norm in the primary dimension is $\Gg^2_1=0.16$, and standard deviation is $\Gg_1=0.4$, as in the dashed reaction norm of \Fig{multipeak}a. The standard deviation in the secondary dimension is $\Gg_2=0.01\Gg_1$.  The smoothed fitness surface rises steadily to a peak along its ridge in the primary dimension, tracing the same path as the dashed curve in \Fig{multipeak}c.  (c and d) Increasingly broad reaction norms in the secondary dimension with standard deviations of $0.1\Gg_1$ and $\Gg_1$, respectively. The baseline truncates small fitness values, which are considered inviable.}
\label{fig:dimension}
\end{figure}

To visualize the multidimensional problem, begin with the one-dimensional fitness landscape in \Fig{multipeak}b.  Now consider two phenotypic dimensions.  Assume that fitness concentrates along one dimension, as in \Fig{dimension}a.  In that plot, only a narrow band of phenotypes along the second phenotypic dimension produces viable individuals.  In the first dimension, fitness rises and falls along the same peaks and valleys as in \Fig{multipeak}b. Thus, both figures show essentially the same fitness landscape, but in the second case the nearly one-dimensional landscape is embedded in a second dimension (fitnesses scale logarithmically in \Fig{dimension}).

In two dimensions, the reaction norm will smooth phenotypes along both trait axes.  When the reaction norm varies mostly along the same dimension as the variation in fitness, as in \Fig{dimension}b, then we obtain the same smoothing as in one dimension (dashed curve of \Fig{multipeak}c).  When the reaction norm varies in both directions, as in \Fig{dimension}d, then the smoothed surface has very low fitness even at its peak.  The low fitness occurs because the randomly generated reaction norm produces phenotypes spread across two dimensions.  Most of those phenotypes fall off of the one dimensional concentration of fit phenotypes.  

\begin{figure}[t]
\includegraphics[width=0.8\hsize]{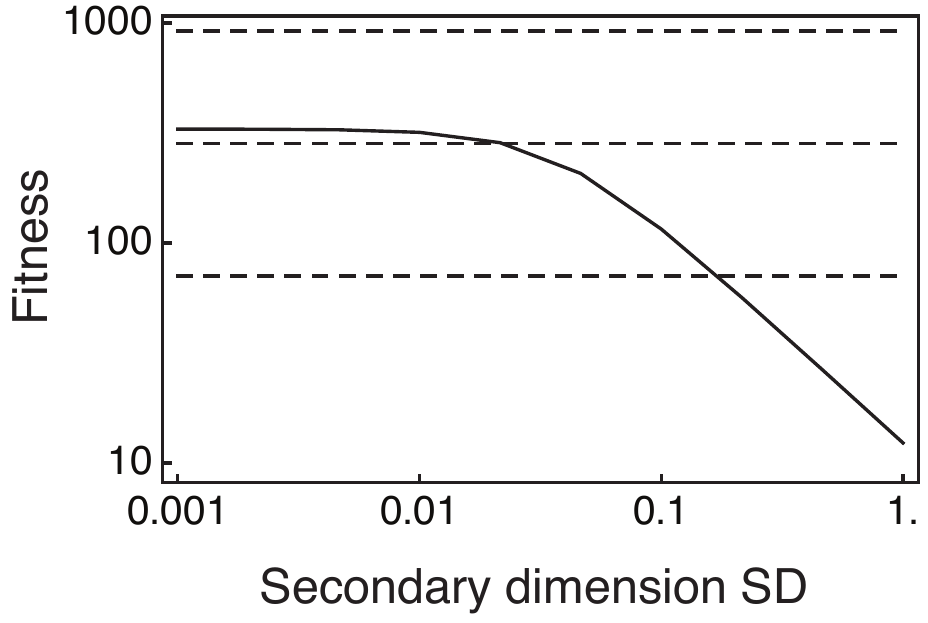}
\caption{Decline in fitness with an increasingly broad reaction norm away from the primary concentration of fitness.  The dashed lines show the fitness associated with the high, medium and low peaks of the fitness landscape in \Fig{dimension}a.  The solid curve shows the highest point of the fitness functions smoothed by the reaction norms of \Fig{dimension}b--d, with standard deviation of the reaction norm increasing in the secondary dimension.  The secondary dimension standard deviation value shown in the plot gives the amount by which the primary dimension standard deviation is multiplied in the second dimension.  The multipliers $0.01, 0.1, 1.0$ correspond to the three smoothed fitness surfaces in \Fig{dimension}b--d.  When the secondary dimension is narrow, for example, reduced in width by a factor $0.01$, then the smoothed fitness peak is higher than the intermediate fitness peak of the unsmoothed landscape, as in \Fig{multipeak}c. As width in the secondary dimension increases, the cost of exploring in a dimension away from the concentration of fitness causes the peak of the smoothed fitness landscape to drop very low, illustrating the very high cost of exploring in more dimensions than the concentration of fitness.}
\label{fig:dimension2}
\end{figure}

In general, when the dimensionality of the reaction norm exceeds the dimensionality of the fitness concentration, then a random search process is inefficient.  The cost of exploration is so high that even the best average phenotype for a genotype has fitness, $F(\xbar)$, lower than the lowest peak of the fitness landscape, $f(x)$.  Here, $\xbar$ and $x$ represent multidimensional phenotypes. \Fig{dimension2} illustrates the cost of exploration in relation to the spread across dimensions.  

In summary, if the space of possible trait combinations spreads over greater dimensions than the concentration of fitness, then randomly generated variations will produce mostly worthless variants.  The search cost is high, and average performance for a widely spread reaction norm is low.  The smoothed fitness surface may have a steadily rising path to its fitness maximum from many initial points, but the height of the fitness peak is so low that a broad reaction norm will often be strongly selected against.  

The following sections describe two processes that may offset the high cost of developmental variation.  First, the broad search space may be covered by genetic variants,  

\begin{figure}[H]
\begin{minipage}{\hsize}
\parindent=15pt
\noterule
{\bf \noindent\BoxLabel. Recent literature}
\noterule
\textcite{ancel00undermining} analyzed smoothed fitness surfaces and the consequences for evolutionary rate.  She emphasized three important points.  

First, learning accelerates evolution only under certain conditions.  The examples in the text illustrate this point by showing that learning mainly accelerates evolution through discovery of viable phenotypes or in the smoothing of a multipeaked fitness surface.  Otherwise, the smoothing of fitness surfaces may lower the maximum fitness that can be attained, reducing the slope and the evolutionary rate to the peak.  

Second, although learning  may accelerate evolution, it is not necessarily true that learning evolved because it accelerates evolution.  The evolutionary consequence of a trait is distinct from whether or not the trait evolved because of its potential to alter subsequent evolutionary dynamics.  The literature discusses this distinction under the topic of evolvability.  Evolvability has developed into a large subject of its own \citep{wagner96complex,kirschner98evolvability,pigliucci08is-evolvability,rajon11evolution,woods11second-order}.  Holland's \autocite{holland75adaptation} distinction between exploration and exploitation captures aspects of the later developments on evolvability.

Third, Ancel noted historical precedents for the idea that phenotypic variance may eliminate otherwise uncrossable valleys in fitness landscapes \autocite{wright31evolution,lande80genetic,whitlock97founder}. 

A large literature develops issues related to Ancel's three points and the broader problems of how reaction norms affect fitness surfaces.  I list a small sample \autocite{de-jong90quantitative,gavrilets93the-genetics,anderson95learning,frank96the-design,turney96myths,mayley97landscapes,turney97evolution,pigliucci01phenotypic,rice02a-general,hall03baldwin,price03the-role,gavrilets04fitness,mills06on-crossing,pigliucci06phenotypic,crispo07the-baldwin,suzuki07repeated,lande09adaptation,chevin10when,chevin10adaptation,gavrilets10rapid}.

\textcite{west-eberhard03developmental} discusses many empirical issues and examples.  Recent studies in evolutionary biology provide new data or summaries of the literature \autocite{aubret09genetic,bell11behavior}. Articles in microbiology and cancer research have also developed the relation between nonheritable phenotypic variation and evolutionary process \autocite{rubin90the-significance,booth02stress,sumner02phenotypic,yomo05phenotypic,avery06microbial,niepel09non-genetic,spencer09non-genetic,altschuler10cellular,kaneko11characterization}.

In the text, I discuss the synergism between genetic and developmental variation.  I am not aware of literature related to that issue.  However, given the many papers on the general topic, the synergism between genetics and development may have come up previously.
\noterule
\end{minipage}
\end{figure}
\boxlabel{literature2}

\noindent with developmental variation searching only the local regions around each genotypic variant. 

Second, developmental variation may be biased in a way that tends to match the environment.  If a developmental or learning process brings the phenotype close to the concentration of fitness in a high dimensional space, then some additional random variation can greatly increase the rate of adaptation.  In this case, the fitness surface is smoothed to provide a steady path of increasing fitness, and the developmental bias that brings the center of the phenotypic distribution close to the fitness concentration mitigates the large cost of search in high dimensional phenotypic spaces.

\section{Synergism between phenotypic and genetic variation}

\begin{figure*}[t]
\includegraphics[width=0.95\hsize]{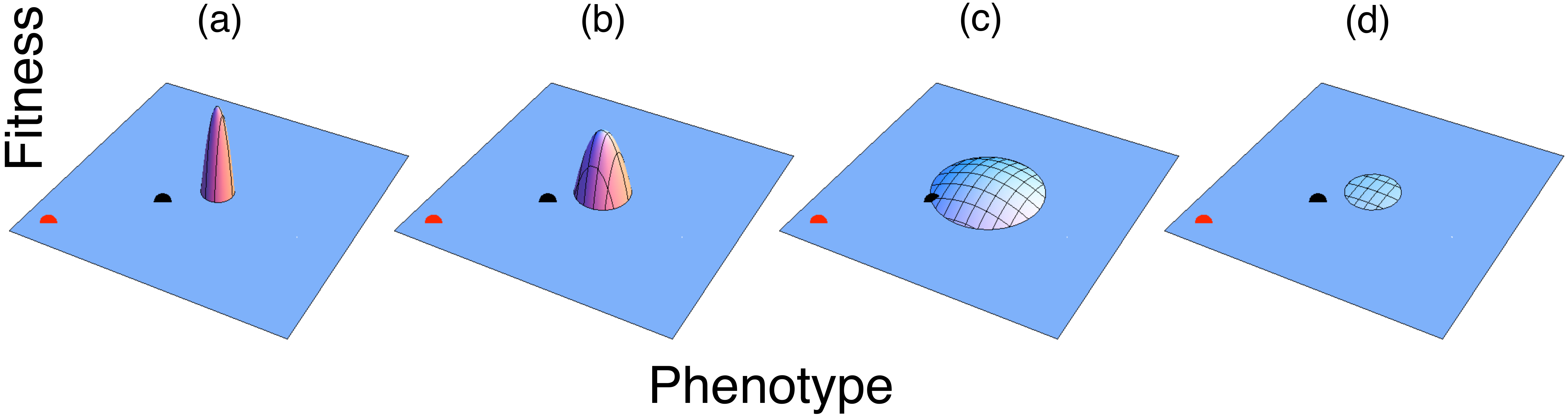}
\caption{Fitness in two phenotypic dimensions after challenge by a novel or extreme environment. The peak corresponds to the favored phenotype after environmental challenge.  The black dot shows the average phenotype for a genotype near the new fitness peak. The red dot shows the average phenotype for a genotype relatively far from the new peak.  A sufficiently broad reaction norm allows the nearby genotype to survive, providing an opportunity for natural selection to drive the population up the smoothly increasing path to the new fitness peak. By contrast, the distant genotype cannot survive the environmental challenge.  (a) The fitness landscape, $f(x)$, showing the direct relation between phenotype and fitness when not smoothed by a reaction norm.  (b) A relatively narrow reaction norm smooths the fitness peak, $F(\xbar)$, but not sufficiently to allow the nearby genotype to survive. (c) A broader reaction norm allows the nearby genotype to survive, with subsequent opportunity for natural selection to drive the population to the peak. (d) An increasingly broad reaction norm causes the smoothed fitness peak to sink mostly below the fitness truncation level, so that the nearby genotype cannot survive.  All plots show a bivariate normal fitness surface with mean $(1/2,1/2)$ and variance $\Gs^2+\Gg^2$, with $\Gs^2=0.01$ for the fitness landscape, and $\Gg^2$ for the reaction norm of $0,0.01, 0.07,0.11$ for plots left to right.  The heights are the natural logarithm of fitness, with a truncation base of $\log(10)$.  The nearby black dot is at $(17/40,17/40)$, and the far red dot is at $(11/40,11/40)$.}
\label{fig:synergy}
\end{figure*}

A broad reaction norm may enhance survival and subsequent opportunity for improved fitness.  But those benefits arise only when a genotype is sufficiently close to a fitness peak.  \Fig{synergy} illustrates the problem.  

\Fig{synergy}a shows the fitness peak in a novel environment.  The dots show the locations of alternative genotypes, placed by their average phenotypes in two dimensions.  Neither genotype has positive fitness.  Both will die out.  In that plot, the fitness peak is the direct fitness landscape, unsmoothed by a reaction norm.  In \Fig{synergy}b, the reaction norm is relatively narrow, smoothing the fitness landscape. But that smoothing is not enough to place either genotype on the nonzero fitness surface.  Both genotypes still die out.  

The broader reaction norm in \Fig{synergy}c smooths the fitness surface more widely.  That additional smoothing allows the nearby genotype to survive.  Subsequent small genetic variations would allow natural selection to drive the surviving population up the path of increasing fitness to the fitness peak.  

The distant (red) genotype cannot survive even with the broad reaction norm of \Fig{synergy}c.  The contrast between the nearby and distant genotypes emphasizes a key point.  A genotype must be sufficiently close to the nonzero part of the smoothed fitness surface in order for the developmental variation of the reaction norm to allow survival---the touching of the fitness surface.  If a genotype touches the fitness surface, then it can seed a population in which small genetic variations allow subsequent adaptation by climbing the surface to the peak.

In a high dimensional space, any single genotype is unlikely to be located sufficiently close to a fitness peak after a significant change in the environment or in response to an unpredictable challenge.  Synergism between genetic variation and the phenotypic variation of reaction norms provides one solution to this search problem.

\begin{figure*}[t]
\includegraphics[width=0.95\hsize]{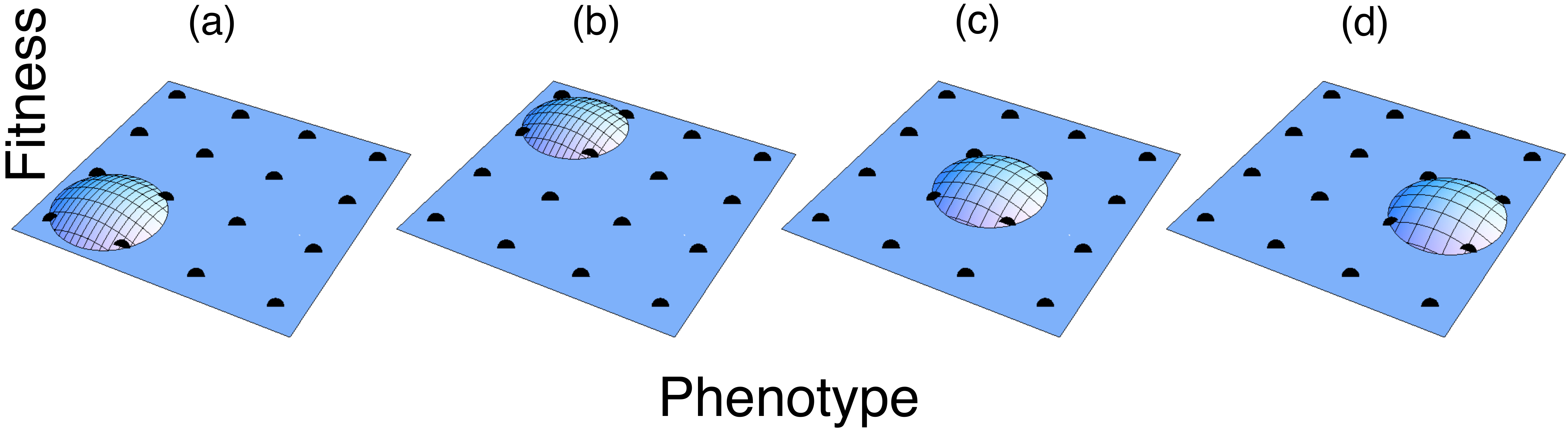}
\caption{Synergism between the reaction norm and genetic variation allows rapid adaptation to novel or extreme environments.  In this case, a population has multiple genotypes, each genotype located at one of the dots.  The smoothed fitness surface is sufficiently broad that, for any location of the fitness peak after environmental challenge, the fitness surface touches at least one of the genotypes.  The genotypes that touch the fitness surface survive, allowing the surviving population the potential subsequently to evolve up the fitness surface to the new peak.  The dots are located at all bivariate pairs from $\{(5+6i)/40,(5+6j)/40\}$ for $i,j=1,\ldots,4$.  The fitness surface has variance $\Gs^2+\Gg^2=0.08$, as in \Fig{synergy}c.}
\label{fig:synergy2}
\end{figure*}

\Fig{synergy2} illustrates synergism between genetic and phenotypic variation.  The dots represent different genotypes.  Each genotype has a different combination of average phenotypic values in two dimensions.  The array of dots shows the genetic diversity in the population.  The smoothed fitness surface has the same fitness peak and reaction norm shape as in \Fig{synergy}c.  In \Fig{synergy2}, the location of the fitness surface varies in the different plots, illustrating different environmental challenges.  No matter where the newly favored fitness surface arises upon environmental challenge, the genetic diversity in the population provides at least one genotype on the nonzero part of the novel fitness surface.  Those genotypes on the surface can survive the novel challenge.  Subsequent small genetic variations around a surviving genotype allow the population to evolve up the fitness surface to the peak set by the novel environmental challenge.  

Synergism between genetic and phenotypic variation divides the adaptive search problem into three parts.  Genetic variation covers widely separated locations in the phenotype space.  Reaction norms cover the phenotype space around each genotype.  Any genotype on a nonzero part of a novel fitness surface can survive and subsequently adapt by small genetic variations and natural 

\begin{figure}[H]
\begin{minipage}{\hsize}
\parindent=15pt
\noterule
{\bf \noindent\BoxLabel. Vertebrate immunity}
\noterule
Invading pathogens present a vast diversity of foreign molecules that must be recognized. The vertebrate adaptive immune system develops antibodies by synergism between phenotypic and genetic variation, following the general three-part search process described in the text  \autocite{frank02immunology,murphy07janeways}.

First, to generate genetic diversity, B cell lineages within the body undergo programmed genetic recombination early in life. That recombination yields genetically distinct cellular clones. Each clone produces a distinct antibody. 

Second, each antibody type from this initial diversity tends to bind relatively weakly to a variety of foreign antigens. In this regard, the original or ``natural'' antibodies trade the cost of weak binding for the benefit of a phenotypically diverse response---a broad reaction norm.  Upon challenge with a foreign antigen, those B cells with matching antibodies are stimulated to expand clonally. That clonal expansion can be thought of as survival and reproduction of those genotypes that land on the fitness surface imposed by the unpredictable invader.  

Third, the weakly binding antibodies undergo a programmed round of hypermutation to the antibody binding site and selection favoring variants that bind more tightly to the foreign antigen. This affinity maturation produces tightly binding and highly adapted antibodies in response to the novel challenge.  Put another way, the initially stimulated antibodies on the edge of the ``fitness surface'' climb the surface toward the fitness peak.  

In the process of climbing the fitness peak by local genetic variation and natural selection, the refined antibodies match more closely to the environmental challenge.  In particular, the refined antibodies narrow their reaction norm by increasing their binding affinity for close matches and reducing their binding affinity for slightly mismatched binding.  

In summary, the ability of the adaptive immune system to respond to the huge diversity of potential challenges depends on its synergism between genetic variability and the reaction norm. The initial natural antibodies arise from genetically diverse clones produced by recombination. That genetic diversity by itself could not cover the huge space of possible challenges. The broad reaction norm around each genetic variant allows protection against novel challenge. Once partial recognition is achieved through the natural antibodies, the system refines the match locally by affinity maturation.
\noterule
\end{minipage}
\end{figure}
\boxlabel{immunity}

\noindent selection.  See \Boxx{immunity} for an example of the synergism between genetic and nonheritable phenotypic variation.

\section{Matching the environment by plasticity or learning}

To survive a novel environmental challenge, a phenotype must be near the nonzero part of the new fitness landscape.  A population may survive by having a variety of genotypes that produce different phenotypes, increasing the chance that at least one of the phenotypes will be close to a new fitness peak.  Alternatively, a single genotype may be able to produce diverse phenotypes by matching phenotypic expression to the particular environment.  The developmental flexibility to match environments may arise by phenotypic plasticity or learning.  

Plasticity or learning may not be able to match exactly a novel or extreme environmental challenge.  But if a developmental response to the environment can move the phenotype sufficiently close to the nonzero part of the new fitness landscape, then the genotype may survive and subsequently adapt \autocite{baldwin96a-new-factor,waddington42canalization,west-eberhard03developmental}.  Developmental flexibility is simply another process that alters the shape of the fitness surface.

The adaptive search problem has three phases, similar to the three aspects of search described in the prior section.  First, partially matching expression to the environment brings the phenotype close to the new fitness landscape.  Second, random perturbations of phenotype occur around the location set by the process of environmental matching.  Third, any genotype on a novel fitness surface can survive and subsequently adapt by small genetic variations and natural selection.  

\begin{figure}[t]
\includegraphics[width=0.8\hsize]{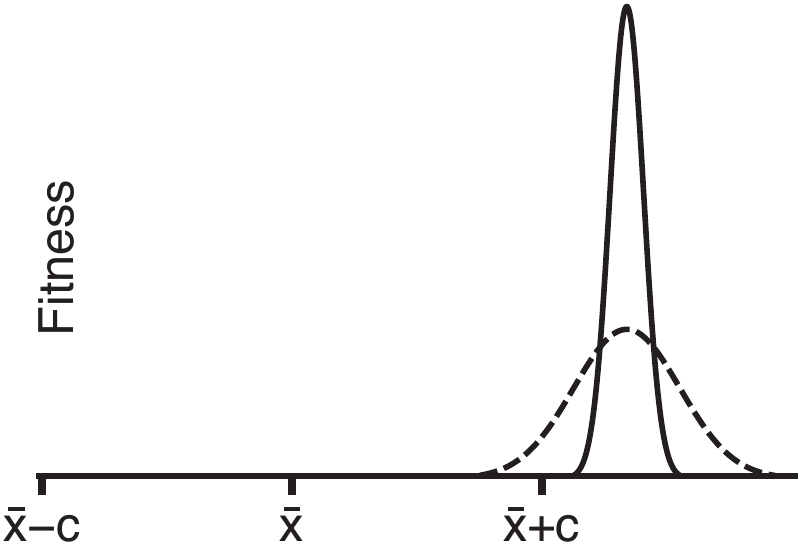}
\caption{Plasticity or learning provides a partial match of phenotype to novel or extreme environmental challenge.  A genotype's default expression has average phenotypic value $\xbar$.  An individual can modify average trait expression in response to the environment.  The average expressed phenotype can be any value in the range $\xbar\pm c$.  The environmental challenge defines the fitness landscape that relates phenotype to fitness, here shown as the solid peak.  Phenotypic expression, modulated by a match to the environment, shifts the phenotype to $\xbar+c$.  However, the fitness associated with a phenotype of $\xbar+c$ is zero, because that value remains outside the range of viable phenotypes.  Suppose $\xbar+c$ is the average phenotype expressed, and random perturbations of expression cause variability in phenotype around that average value.  The random component of phenotype smooths the fitness landscape, leading to the dashed fitness surface.  The expressed average phenotype, $\xbar+c$, now falls within the smoothed fitness surface, allowing the genotype to survive.  Subsequent adaptation may allow improved fitness, by altering the range of phenotypes that can be expressed so that a match to the fitness peak may be achieved. }
\label{fig:plasticity}
\end{figure}

\Fig{plasticity} illustrates the three aspects of adaptive search.  Suppose a genotype expresses phenotypes centered at $\xbar$.  In the first aspect of adaptive search, a genotype can adjust phenotypic expression to match the environment.  The possible range of phenotypes varies from $\xbar-c$ to $\xbar+c$.  The phenotype expressed by environmental matching is the new average value, around which random perturbations may occur.  In the figure, the solid peak shows the fitness landscape imposed by a novel or extreme environmental challenge.  The example genotype can come close to the new peak by modulating expression to produce an average phenotype of $\xbar+c$.  However, if no random variation occurs around $\xbar+c$, that phenotype falls outside the range of phenotypic values that can survive.

In the second aspect of adaptive search, the genotype may produce phenotypes randomly distributed around the mean value of $\xbar+c$.  Those random fluctuations smooth the fitness landscape, shown by the dashed curve.  The average phenotype $\xbar+c$ can now survive.  Matching the environment allowed expression of mean phenotype $\xbar+c$, and random fluctuations in phenotype smoothed the nearby landscape sufficiently.  

Once the genotype achieves survival, the third phase of adaptation may proceed.  In this case, the mean phenotype $\xbar+c$ has low fitness on the dashed fitness surface.  But the fitness surface has a smoothly increasing path to the peak of maximum fitness.  Genetic variations in the genotype may shift the range of phenotypes that can be produced, allowing natural selection to drive the population up the fitness surface to the peak.

\begin{figure}[H]
\begin{minipage}{\hsize}
\parindent=15pt
\noterule
{\bf \noindent\BoxLabel. A simple prediction}
\noterule
Frequent exposure to novel or extreme environmental challenge favors a greater capacity for evolutionary response.  Faster evolutionary response may be achieved by enhanced generation of genetic variability.  For example, in experimental evolution studies, strong competition favors strains of bacteria with a high mutation rate \autocite{loh10optimization}.  Strong competition can be thought of as a form of extreme challenge.  However, interference between genetically distinct competing clones complicates the association between mutation rate and evolutionary rate \autocite{rainey99evolutionary}.

Faster evolutionary response may also be achieved by enhanced phenotypic variability through learning or plasticity.  However, it may be difficult to test empirically whether learning or plasticity are direct responses to increased environmental challenge.  

The problem is that learning and plasticity tend to be complex adaptations.  To build those complex adaptations takes a long time and many changes.  That slow process means that one would have to measure the change in evolutionary pressure over the long period during which learning or plasticity evolve.  Ultimately, one would need direct measures of environmental pressure in periods of weak challenge and in periods of strong challenge, and measures of the time course of change in learning or plasticity.  That is hard to do.

A simpler prediction concerns the rapid increase in the reaction norm in response to an environmental challenge.  If a population encounters a novel or extreme environmental challenge, it may respond by broadening its reaction norm.  A broader reaction norm can be achieved simply by increasing the tendency for stochastic perturbations during development---a breakdown in the normal homeostatic processes that keep phenotypes within narrow bounds.  This prediction about the broadening of the reaction norm follows immediately from Holland's \autocite{holland75adaptation} contrast between exploitation and exploration and the quote given in the text by \textcite{haldane32the-causes}. Many specific models analyze how the reaction norm responds to environmental challenge.  \textcite{lande09adaptation} provided new models and a good overview of the literature.

In this case, one simply needs to measure the change in environmental pressure over short periods of time and the associated change in the reaction norm of the population.  It would also be interesting to measure how a broader reaction norm acts synergistically with genetic variation to speed the evolutionary response.
\noterule
\end{minipage}
\end{figure}
\boxlabel{predict}

\section{Conclusion}

Evolutionary theory emphasizes genetic variation as the source of evolutionary novelty.  By the standard theory, the usual sequence would be a novel environmental challenge, genetic variation either already present or arising de novo, and evolutionary response to the novel environment by change in gene frequency.  

In this classical evolutionary theory, genetics provides the source of phenotypic variation on which natural selection acts.  By contrast, development may generate the phenotypic novelty that initiates adaptation to environmental challenge.  The sequence would be novel environmental challenge, initial survival by those individuals with a phenotypic norm of reaction that overlaps the new fitness surface, and subsequent adaptation by genetic variants from those phenotypes that survive the initial challenge.  

\textcite{west-eberhard03developmental} traces the theoretical foundations of this topic from the late 19th century.  Since that time, the idea that developmental processes may play a key role in initiating adaptation has never been popular.  Evolutionary change is usually tied in thought to genetic change. Nonheritable phenotypic variation by itself is therefore usually believed not to accelerate evolutionary rate.  

The original theories of learning, developmental plasticity, and reaction norms have always understood the relations between genotype, phenotype, environment, and evolutionary change.  However, the jargon from those theories is thick: the Baldwin effect, genetic assimilation, reaction norms, hopeful monsters, niche construction, and environmentally induced evolution.  Each variant theory invoked special environmental conditions, developmental processes, and interactions with genetics.  And each in its own way jousted with the ghost of Lamarck.  A casual observer could be forgiven for steering clear of the whole mess.  Wisdom suggested to wait for clear empirical examples.  Induction still dominates mainstream thought in biology.  

Many years ago, I read Hinton and Nowlan's \autocite{hinton87how-learning} article and Maynard Smith's \autocite{maynard-smith87when} related essay on the Baldwin effect.  They focused on the essential theoretical point.  Learning smooths the fitness surface, changing evolutionary dynamics in a way that greatly accelerates adaptation to novel or extreme environmental challenges.  When one views the whole confusing field in that simple light, one sees that all the complexities of the theories and mechanistic details of phenotypic variability ultimately reduce to the same point. Developmental variation smooths the fitness landscape.  A smoothed fitness landscape profoundly alters evolutionary dynamics, particularly in response to novel or extreme environmental challenge.

\begin{acknowledgments}
Marsha Rosner's insight into cellular variability and the evolutionary dynamics of cancer stimulated this work. My research is supported by National Science Foundation grant EF-0822399, National Institute of General Medical Sciences MIDAS Program grant U01-GM-76499, and a grant from the James S.~McDonnell Foundation.  I completed this article while supported by a Hogge-Baer Visiting Professorship in Cancer Research at the University of Chicago. 
\end{acknowledgments}

\bibliography{main}

\end{document}